\documentclass[aps,superscriptaddress,eqsecnum,nofootinbib,preprintnumbers]{revtex4}

\usepackage{amsmath,amssymb,graphicx,amsfonts}
\usepackage{mathrsfs}
\usepackage{comment}
\usepackage{xcolor}
\usepackage[shortlabels]{enumitem}
\usepackage[unicode, colorlinks=true, linkcolor=red, citecolor=blue, filecolor=blue, urlcolor=blue, linktocpage, breaklinks]{hyperref}
\usepackage{url}

\begin{document}
\title{Geodesics and gravitational waves in chaotic extreme-mass-ratio inspirals: \\the curious case of Zipoy-Voorhees black-hole mimickers}

\author{Kyriakos Destounis*}
\affiliation{Dipartimento di Fisica, Sapienza Università di Roma, Piazzale Aldo Moro 5, 00185, Roma, Italy}
\affiliation{INFN, Sezione di Roma, Piazzale Aldo Moro 2, 00185, Roma, Italy}
\affiliation{Theoretical Astrophysics, IAAT, University of T{\"u}bingen, 72076 T{\"u}bingen, Germany}
\email{kyriakos.destounis@uniroma1.it}
\author{Giulia Huez}
\affiliation{Theoretical Astrophysics, IAAT, University of T{\"u}bingen, 72076 T{\"u}bingen, Germany}
\affiliation{Physics Department, University of Trento, Via Sommarive 14, 38123 Trento, Italy}
\author{and Kostas D. Kokkotas}
\affiliation{Theoretical Astrophysics, IAAT, University of T{\"u}bingen, 72076 T{\"u}bingen, Germany} 
\affiliation{Section of Astrophysics, Astronomy, and Mechanics, Department of Physics, National and Kapodistrian University of Athens, Panepistimiopolis Zografos GR15783, Athens, Greece}

\begin{abstract}
Due to the growing capacity of gravitational-wave astronomy and black-hole imaging, we will soon be able to emphatically decide if astrophysical dark objects lurking in galactic centers are black holes. Sgr A*, one of the most prolific astronomical radio sources in our galaxy, is the focal point for tests of general relativity. Current mass and spin constraints predict that the central object of the Milky Way is supermassive and slowly rotating, thus can be conservatively modeled as a Schwarzschild black hole. Nevertheless, the well-established presence of accretion disks and astrophysical environments around supermassive compact objects can significantly deform their geometry and complicate their observational scientific yield. Here, we study extreme-mass-ratio binaries comprised of a minuscule secondary object inspiraling onto a supermassive Zipoy-Voorhees compact object; the simplest exact solution of general relativity that describes a static, spheroidal deformation of Schwarzschild spacetime. We examine geodesics of prolate and oblate deformations for generic orbits and reevaluate the non-integrability of Zipoy-Voorhees spacetime through the existence of resonant islands in the orbital phase space. By including radiation loss with post-Newtonian techniques, we evolve stellar-mass secondary objects around a supermassive Zipoy-Voorhees primary and find clear imprints of non-integrability in these systems. The peculiar structure of the primary, allows for, not only typical single crossings of transient resonant islands, that are well-known for non-Kerr objects, but also inspirals that transverse through several islands, in a brief period of time, that lead to multiple glitches in the gravitational-wave frequency evolution of the binary. The detectability of glitches with future spaceborne detectors can, therefore, narrow down the parameter space of exotic solutions that, otherwise, can cast identical shadows with black holes.
\end{abstract}
	
\maketitle

\section{Introduction}

The Schwarzschild spacetime \cite{Schwarzschild:1916uq} is unequivocally the simplest and most remarkable black hole (BH) solution of general relativity. It describes a vacuum compact object with an event horizon and a static, spherically-symmetric exterior. The remarkable symmetry properties of Schwarzschild geometry places it at the top of Einstein field equation solutions in what regards its simplicity and singular externally-observable property; the gravitational mass. Its ultimate successor, the Kerr metric \cite{Kerr:1963ud}, is undoubtedly the most successful and astrophysically-relevant solution of the vacuum field equations that describes a spinning, stationary and axisymmetric BH with an oblate, spheroidal external geometry. Despite the fact that Kerr BHs possess less symmetries than Schwarzschild spacetimes, they compensate by including spin; a very crucial aspect of most astrophysical compact objects, and further form an integrable (separable) system of equations of motion for massless and massive particles due to the existence of the Carter constant \cite{Carter:1968rr}. The above statement of integrability translates to the absence of chaos in geodesics around Kerr BHs, and is traced trivially to Schwarzschild spacetime \cite{Contopoulos_book}.

From an astrophysical perspective, a significant volume that surrounds BHs in the Universe consists of plasma, accretion disks, matter configurations and halos that extend significantly far away. Thus, it is hard for one to realize BHs in pure vacuum, especially those occupying the centers of galaxies, where a perplex and highly dynamical environment is present. The historic observations of shadows from supermassive compact objects in the center of M87* galaxy \cite{EventHorizonTelescope:2019dse} and Sgr A* in our galaxy \cite{EventHorizonTelescope:2022wkp} has opened a new realm of observational yield with BH imaging. An accretion disk can, in principle, deform the surrounding geometry of a BH so that it continuously deviates from a Schwarzschild or Kerr description, causing degeneracies between a multitude of other exact, though more exotic, solutions that may mimic the observed shadow of supermassive compact objects. 

To study potential degeneracies present in current electromagnetic observations, the literature usually operates in spacetime deviations from the Kerr description, known as bumpy/parameterized \cite{Collins:2004ex,Glampedakis:2005cf,Vigeland:2011ji,Johannsen:2011dh,Emparan:2014pra,Cardoso:2014rha,Rezzolla:2014mua,Konoplya:2016jvv,Moore:2017lxy} or non-Kerr compact objects \cite{Sato:1972,Kinnersley:1978,Manko_1992,Manko:2000ud,Manko:2000sg,Bambi2017,Destounis:2020kss}. A distinctive feature of some of these objects is the absence of Carter symmetry, which leads to chaotic phenomena in particle-dynamics. In what regards Sgr A*, though, spin is not as crucial as the deviation from spherical symmetry itself, since current constraints on its spin conclude that it is less than $10\%$ of its maximal allowed \cite{Melia:2001fp,Fragione:2020khu}. Therefore, a plethora of Sgr A*-like objects in the Universe can be sufficiently modeled as Schwarzschild (or slowly-rotating Kerr) BHs or exotic BH mimickers.

Gravitational-wave (GW) astrophysics has been proven to be an extraordinary tool to break the aforementioned degeneracies, thus synergies between GW and shadow observations are necessary in order to search for the ultimate spacetime description of known astrophysical compact objects. To that end, the LIGO/Virgo/Kagra collaboration has been flooding our databases with numerous GW detections from coalescing compact objects \cite{LIGOScientific:2021djp}. Such detectors, although extremely successful so far (at the level of groundbreaking), have the unfortunate attribute of being plagued by planetary noise, since they are placed on Earth, and present unavoidable limitations due to their scale, that play a crucial role in their target span and precision.

The Laser Interferometer Space Antenna (LISA) \cite{LISA:2017pwj} is a space-borne GW detector that will open new realms in GW astrophysics, due to its unprecedented level of accuracy \cite{Baibhav:2019rsa,Amaro-Seoane:2022rxf,LISA:2022kgy,Karnesis:2022vdp}. It will target, in particular, mHz sources of GWs which are undetectable with current ground-based detectors. One of the prime objectives of LISA (and other space-based programs \cite{TianQin:2015yph,Ruan:2018tsw,Ruan:2020smc}) is the detection of GWs from extreme-mass-ratio inspirals (EMRIs) \cite{Glampedakis:2005hs,Gair:2017ynp}, which involve a primary supermassive compact object, such as those lurking in galactic cores, and a secondary stellar-mass compact object. We expect that the detection of EMRIs will allow for exquisite measurements of EMRI parameters, the multipolar structure of the primary of the EMRI \cite{Ryan:1995wh,Ryan:1997hg,Ryan:1997kh,Pacilio:2020jza,Vaglio:2022flq,Vaglio:2023lrd}, and, in general, very accurate tests of GR \cite{Babak:2017tow}. Currently, the most significant tool of perturbations theory to study EMRI evolution is the gravitational self-force \cite{Poisson:2004gg,Barack:2018yvs}, though it has been proven to be a daunting task to undertake and has only been calculated up to second order till date \cite{vandeMeent:2017bcc,Isoyama:2021jjd}. However, other techniques, such as those that will be used in this work, has proven accurate enough to capture the dissipative effects of EMRI evolution, when compared to Teukolsky-based waveforms \cite{Babak:2006uv}, which utilize weak field post-Newtonian techniques to find the fluxes together with general-relativistic kludge schemes to evolve the EMRI accurately enough \cite{Glampedakis:2002cb,Glampedakis:2002ya,Glampedakis:2005hs} even when the primary spacetime of the EMRI is quasi-Kerr in nature \cite{Gair:2005ih,Barack:2006pq,Glampedakis:2005cf,Gair:2007kr,Lukes-Gerakopoulos:2010ipp,Destounis:2020kss,Destounis:2021mqv,Destounis:2021rko}. Environmental effects and spacetime deformations in EMRIs should, therefore, be taken into account in order to maximize the scientific yield from these sources \cite{Chakrabarti:1995dw,Ryan:1995wh,Ryan:1997hg,Ryan:1997kh,Barausse:2006vt,Barausse:2007dy,Eda:2013gg,Macedo:2013qea,Barausse:2014tra, Cardoso:2016olt,Cardoso:2019rou,Toubiana:2020drf,Caputo:2020irr,Cardoso:2021sip,Zwick:2021dlg,Zwick:2022dih,Speri:2022upm,Sberna:2022qbn,Polcar:2022bwv,Vicente:2022ivh,Speeney:2022ryg,Cardoso:2021wlq,Cardoso:2022whc,Cheung:2021bol,Destounis:2022obl}. Fortunately or not, it is atypical to find integrable spacetimes, especially when complex astrophysical environments are involved or the primary is an exotic compact object \cite{Cardoso:2019rvt}. To that end, a significant spacetime degeneracy is present in Sgr A* and M87*. Current investigations have concluded that an abundance of exotic objects can cast shadows that are practically indistinguishable (in specified regions of their parameter space) from those of Schwarzschild (or Kerr) BHs \cite{Vincent:2015xta,Olivares:2018abq,Abdikamalov:2019ztb,Wielgus:2020uqz,Herdeiro:2021lwl,Wielgus:2021peu,Rosa:2022tfv,Rosa:2022toh}, therefore assuming that Sgr A* and M87* are BHs, only from their shadow silhouette, can lead to significant misinterpretation. Indeed, we need to invoke geodesics \cite{Boshkayev:2015jaa}, accretion disk analyses \cite{Chowdhury:2011aa} and GW ringdown tests \cite{Kokkotas:1999bd,Berti:2009kk,Cardoso:2016rao,Cardoso:2016oxy,Abedi:2016hgu,Maggio:2019zyv,Maggio:2020jml,Abedi:2020ujo,Maggio2020,Vlachos:2021weq,Chatzifotis:2021pak,Chatzifotis:2021hpg,Boyanov:2022ark} in order to understand if supermassive compact objects are typical BHs or exotic in nature.

In this study, we will investigate the simplest deformation of Schwarzschild geometry, the Zipoy-Voorhees (ZV) metric \cite{Zipoy:1966,Voorhees:1970ywo} (also known as the $\gamma$-metric, that has also been generalized in \cite{Gurtug:2021noy,Halilsoy:2022pzr} to include electric charge), that describes a vacuum, static, and spheroidal solution of Einstein equations which is continuously connected to Schwarzschild by a deformation parameter $\delta$ (in our convention). The ZV metric, presenting a spheroidal deformation of an otherwise static geometry can pose as a good model for a static compact object surrounded by a compact environment or an accretion disk, such as those residing in galactic centers. Recent shadow investigations \cite{Abdikamalov:2019ztb} have shown that when the deformation parameter $\delta>1$ then very precise measurements will be needed in order to rule out an exotic compact object described by this geometry. A peculiarity of this solution is the appearance of a curvature singularity at its surface, thus characterizing it as a naked singularity. Another important feature for our analysis is that the ZV metric has a non-zero mass quadrupole and non-integrable geodesics \cite{Kruglikov:2011ky,Lukes-Gerakopoulos:2012qpc,Maciejewski:2013ncb}, despite earlier claims of integrability \cite{Sota:1995ms,Brink:2008xy}. Since non-integrable EMRIs present very distinctive characteristics in phase space, such as prolonged resonant islands where geodesics share the same rational ratio of orbital frequencies \cite{Apostolatos:2009vu,Lukes-Gerakopoulos:2010ipp,Lukes-Gerakopoulos:2014dpa,Destounis:2020kss,Deich:2022vna,Chen:2022znf} and discontinuities in the GW frequencies during island crossings (`glitches') \cite{Destounis:2021mqv,Destounis:2021rko}, here we examine if similar effects are present in ZV EMRIs, and in particular GW glitches from crossings of subdominant resonant islands.

We confirm that the conclusions of Refs. \cite{Kruglikov:2011ky,Lukes-Gerakopoulos:2012qpc,Maciejewski:2013ncb} are correct, that is the ZV metric is non-integrable due to the existence of chaotic layers of plunging geodesics and a series of resonant islands of stability. We further choose a primary supermassive compact object described by the ZV metric and evolve EMRIs with stellar-mass secondaries (with fixed mass ratio) to assess the effect of non-integrability at the orbital and waveform level. We find plateaus in the dissipative evolution of the ratio of radial and polar frequencies, that designate the crossing of resonant islands, and subsequently observe glitches in the GW frequency evolution of the EMRI. Due to the atypical structure of the ZV primary, a variety of successive resonant islands accumulate close to the edge of bound geodesics. By evolving EMRIs through successive resonances, with generic initial conditions, we find that consecutive glitches can appear in short timescales (of order of several days) in the GW frequency evolution of the inspiral. Since typical glitches experienced by non-Kerr EMRIs, with the same mass ratio as the ZV one, are usually separated by months or even years of dissipative evolution \cite{Destounis:2021mqv,Destounis:2021rko}, the detection of multiple GW glitches in brief periods of time may demonstrate that very slowly-rotating supermassive compact objects are not Schwarzschild (or slowly-rotating Kerr) BHs. These findings will contribute in placing tighter constraints on exotic geometries, such as naked singularities, and narrow down the parameter space of viable BH mimicker primaries that can imitate the shadow of supermassive BHs. 

In what follows we use geometrized units so that the gravitational constant and speed of light satisfy $G=c=1$.

\section{the Zipoy-Voorhees metric}

The ZV spacetime \cite{Zipoy:1966,Voorhees:1970ywo} describes a two parameter family of exact vacuum solutions to the Einstein equations that are static, axisymmetric and asymptotically flat. The line element in Erez-Rosen coordinates \cite{Erez_1959,Esposito:1975qkv} reads
\begin{align}\label{ZV_metric}
	ds^2=&-F(r) dt^2+F^{-1}(r)\left[G(r,\theta)dr^2+H(r,\theta)d\theta^2+(r^2-2kr)\sin^2\theta d\phi^2\right],
\end{align}
with $k=M/\delta$, where $M$ is the gravitational mass of the object and $\delta$ the deformation parameter, while the functions involved in the metric tensor components are
\begin{align}\label{metric_comp}
	F(r)&=\left(1-\frac{2k}{r}\right)^\delta,\nonumber\\
	G(r,\theta)&=\left(\frac{r^2-2kr}{r^2-2kr+k^2\sin^2\theta}\right)^{\delta^2-1},\\
	H(r,\theta)&=\frac{\left(r^2-2kr\right)^{\delta^2}}{\left(r^2-2kr+k^2\sin^2\theta\right)^{\delta^2-1}}\nonumber.
\end{align}
From Eqs. \eqref{ZV_metric}, \eqref{metric_comp}, it is straightforward to obtain the Schwarzschild limit when $\delta=1$. Since at the Schwarzschild limit, $k=M$, $\delta$ can be interpreted as a measure for how much more (or less) mass $M=k\delta$ the ZV object has when compared to a Schwarzschild BH. Subsequently, the deformation $\delta$ captures the oblateness of the compact object. When $\delta>1$ the ZV geometry describes a spacetime around the central object that is more oblate than a Schwarzschild BH, while when $0<\delta<1$ the central object is more prolate. For $\delta=0$ (which identically means $M=0$) we obtain the Minkowski spacetime.

According to the no-hair theorem, when $0<\delta\neq 1$ holds the event horizon is broken; a true curvature singularity appear at $r=2k$ \cite{Virbhadra:1996cz}, besides the typical one at $r=0$, and the ZV metric describes a naked singularity \cite{Papadopoulos:1981wr}. Specifically, when writing the solution in Weyl coordinates $(\rho,z)$, which are associated with the prolate spheroidal coordinates $(x,y)$ that many pathological solution of GR are written in (see e.g. the Manko-Novikov solution \cite{Manko_1992}), then it has been shown that the curvature singularity naturally appears along the line segment $\rho=0$, if $|z|<k$. For more information regarding the nature of the ZV naked singularity see \cite{Papadopoulos:1981wr,Herrera:1998eq,Kodama:2003ch,Gibbons:2004au}. Interestingly, when spherical symmetry is broken the geometry obtains a non-zero quadrupole moment $M_2=\delta(1-\delta^2)M^3/3$ \cite{Geroch:1970cd,Hansen:1974zz} which eventually leads to the Carter constant (or any other higher order Killing tensor) being broken \cite{Kruglikov:2011ky,Lukes-Gerakopoulos:2012qpc,Maciejewski:2013ncb}.

\section{Orbital dynamics}

Regardless of its peculiar causal structure, the ZV primary we will focus on should possess as many `good' features as those of astrophysical compact objects. It has been shown that the line element \eqref{ZV_metric} has an innermost stable circular orbit (ISCO) when $\delta>1/\sqrt{5}$ at \cite{Chowdhury:2011aa,Lukes-Gerakopoulos:2012qpc}
\begin{equation}\label{ISCO}
	r_\text{ISCO}=k\left(1+3\delta+\sqrt{5\delta^2-1}\right).
\end{equation}
Specifically, the typical ISCO of Schwarzschild BHs that is a special stable, circular and equatorial orbit of the separatrix manifold, that separates between plunge and bound motion, changes structure when $\delta\neq1$ and asymptotic manifolds emanate from an unstable orbit, called the Lyapunov orbit \cite{Contopoulos_1990,Contopoulos:2012nd}. Orbits that cross the Lyapunov orbit will plunge unless certain $E$ and $L_z$ \cite{Lukes-Gerakopoulos:2012qpc} turn the Lyapunov orbit to the ISCO and the orbit becomes stable. For more information regarding the relation between the separatrix and last stable orbits, as well as other special last stable orbits, we refer the reader to \cite{Stein:2019buj}.
Furthermore, if $\delta\geq1/2$ then the geometry has both an ISCO and a photon sphere (PS), where unstable null geodesics accumulate \cite{Cardoso:2008bp,Cardoso:2017soq}, at \cite{Chowdhury:2011aa}
\begin{equation}\label{photon sphere}
	r_\text{PS}=k(1+2\delta).
\end{equation}
Notice that at the Schwarzschild limit where $\delta=1$, $r_\text{PS}=3M$ and $r_\text{ISCO}=6M$ as expected. Hereafter, we will focus on spacetime deformations that are larger than $1/2$ in order to have an exotic central object with a PS and an ISCO that are fundamental hypersurfaces of astrophysically relevant (exotic) compact objects.

\subsection{Geodesics}

A first-order approximation to EMRI evolution can be accomplished through geodesics of a point-particle of mass $\mu$ which plays the role of the secondary orbiting around the primary supermassive compact object. 

The geodesic equations read
\begin{equation}\label{geodesic}
	\ddot{x}^\kappa+\Gamma^\kappa_{\lambda\nu}\dot{x}^\lambda \dot{x}^\nu=0,
\end{equation}
where $\Gamma^\kappa_{\lambda\nu}$ are the Christoffel symbols of spacetime, $x^\kappa$ is the four-position vector, $\dot{x}^\kappa$ is the four-velocity vector and the overdot denotes differentiation with respect to proper time $\tau$.

Stationary and axisymmetric spacetimes, such as Eq. \eqref{ZV_metric}, possess metric tensor components that are $t$- and $\phi$-independent. Therefore, they admit at least two conserved quantities (due to stationarity and axisymmetry) throughout geodesic evolution, namely the energy $E$ and $z$-component of the orbital angular momentum $L_z$
\begin{equation}\label{ELz}
	E/\mu=F(r)\dot{t}, \quad L_z/\mu=\frac{\left(r^2-2kr\right)\sin^2\theta}{F(r)} \dot{\phi}.
\end{equation}
The $t$- and $\phi$-momenta can be expressed with respect to the conserved quantities and the non-zero metric tensor components. Together with the conservation of the rest mass $\mu$ of the secondary, (preservation of four-velocity) which leads to $g_{\lambda\nu}\dot{x}^\lambda\dot{x}^\nu=-1$, the geodesics of test particles present three constants of motion. Specifically, the conservation of the secondary's four-velocity leads to a constraint for bound orbits
\begin{equation}\label{constraint}
	\dot{r}^2+\frac{H(r,\theta)}{G(r,\theta)}\dot{\theta}^2+V_\text{eff}=0,
\end{equation}
where the effective potential $V_\text{eff}$ has the form
\begin{align}\label{effective_pot}
	V_\text{eff}\equiv \frac{1}{g_{rr}}\left(1+\frac{E^2}{g_{tt}}+\frac{L_z^2}{g_{\phi\phi}}\right)=\frac{F(r)}{G(r,\theta)}\left(1-\frac{E^2}{F(r)}+\frac{F(r)L_z^2}{(r^2-2kr)\sin^2\theta}\right).
\end{align}
The curve defined when $V_\text{eff}=0$. i.e. the curve of zero velocity (CZV), can be used in order to choose proper initial conditions that lead to bound orbits in the external vicinity of the primary.

Bound geodesic motion in integrable systems can, generically, be characterized by three orbital frequencies. These frequencies are associated with the radial rate of transition between the periapsis and apoapsis of the geodesic ($\omega_r$), the rate of longitudinal oscillations through the equatorial plane ($\omega_\theta$) and the revolution around the primary ($\omega_\phi$). Generic trajectories with irrational ratios of orbital frequencies span on two-dimensional tori and fill them completely. To the contrary, when the ratio of two orbital frequencies is a rational number then the geodesic is periodic (or resonant) and returns to its initial position after a number of oscillations defined by the multiplicity of the resonance \cite{Contopoulos_book}. Such orbits are special in the sense that they are not phase-space filling and therefore, can directly affect the evolution of EMRIs when encountered \cite{Flanagan:2010cd,Flanagan:2012kg,Brink:2013nna,Ruangsri:2013hra,vandeMeent:2013sza,vandeMeent:2014raa,Brink:2015roa,Berry:2016bit,Speri:2021psr,Gupta:2022fbe,Apostolatos:2009vu,Lukes-Gerakopoulos:2010ipp,Zelenka:2019nyp,Lukes-Gerakopoulos2020,Mukherjee:2022dju,Destounis:2020kss,Destounis:2021mqv,Destounis:2021rko}.

Regardless of the fact that the ZV metric is non-integrable, when the deformation parameter $\delta$ is not too far from unity \cite{Contopoulos_book} then the majority of generic orbits are still characterized by orbital frequencies and elements in accord to the KAM theorem discussed below. Close to resonances, indirect and direct chaotic phenomena appear due to the non-integrability of spacetime \cite{Lukes-Gerakopoulos:2012qpc}, but we stay close enough to $\delta=1$ so that pure chaos never emanates in a direct manner (see Sec. \ref{chaos} for more details on indirect and pure chaos). Therefore, we are still able to define orbital frequencies and only an almost zero-volume of the parameter space has regions of pure chaos (which we cannot spot in our analysis) where orbital frequencies are ill-defined. This will become more obvious in the following sections where all our imprints of chaos are indirect in nature.

\subsection{Inspirals}

To construct the inspiral trajectory we numerically integrate the coupled system of $r,\,\theta$ equations, after utilizing Eqs. \eqref{ELz}, augmented with post-Newtonian (PN) fluxes for the energy and angular momentum, respectively \cite{Glampedakis:2002ya,Glampedakis:2002cb,Barack:2003fp,Gair:2005ih}. This treatment, though approximate, takes into account the dominant contribution of the secondary's radiative backreaction to the spacetime geometry, at second PN order, and results to an adiabatic evolution of the EMRI through time-dependent shifts onto the energy and $z$-component of angular momentum of the secondary. Since the inspiral evolves very slowly, the orbit is treated, in small timescales, as a geodesic, while for long timescales the trajectory is driven adiabatically through successively damped geodesics. This method, known as the hybrid kludge scheme, has been shown to perform very well when compared to Teukolsky-based Kerr waveforms for EMRIs \cite{Babak:2006uv}. 

Notice that the ZV metric has a non-trivial multipolar structure due to the deformation parameter $\delta$, and in particular a non-zero mass quadrupole tensor, as opposed to that of Schwarzschild BHs. At second PN order, the kludge scheme \cite{Glampedakis:2002cb} involves the mass quadrupole moment $M_2$ (for Kerr), thus to construct a more appropriate inspiral around ZV compact objects we apply a modification to the fluxes (see \cite{Barack:2006pq,Gair:2007kr,Apostolatos:2009vu,Lukes-Gerakopoulos:2010ipp,Destounis:2021mqv,Destounis:2021rko}) in order to include the quadrupole moment $M_2$ of the ZV metric, which represents the effect of $\delta$ on the evolution of $E$, $L_z$, and set the spin parameter to zero. The adiabatic approximation, together with the flux augmentation, employed here has recently been found to provide results qualitatively equivalent to evolutions with instantaneous self-force in non-Kerr electromagnetic analogues, which indicates that the methods we use can in principle describe resonance-crossings with sufficient accuracy in EMRIs \cite{Mukherjee:2022dju}. Nevertheless, more accurate inspirals can be built by directly solving the wave equation resulting from metric perturbations and calculating the GW emission at the object and infinity, though this is a an almost impossible task and out of the scope of the phenomenology we are after in this article.

We assume linear variations of the momenta as in \cite{Canizares:2012is,Destounis:2021mqv,Destounis:2021rko}
\begin{align}
	\label{update_E}
	E_1=\frac{E_0}{\mu}+\bigg< \frac{dE}{dt}\bigg>\bigg|_0 N_r\, T_r,\\
	\label{update_L}
	L_{z,1}=\frac{L_{z,0}}{\mu}+ \bigg< \frac{dL_z}{dt}\bigg>\bigg|_0 N_r\, T_r,
\end{align}
where $E_0,\,L_{z,0}$ are the initial energy and $z$-component of the angular momentum, respectively, and $\langle{dE}/{dt}\rangle|_0,\,\langle{dLz}/{dt}\rangle|_0$ are the radiation fluxes calculated at the beginning of the inspiral, through the equations in \cite{Gair:2005ih,Destounis:2021rko}. $T_r$ is the time that the orbit takes to travel from the periapsis to apoapsis and back, while the fluxes \eqref{update_E} and \eqref{update_L} are updated every $N_r$ cycles for the whole EMRI evolution. 

\subsection{Detecting chaos}\label{chaos}

To understand the phase space structure of orbits around the ZV primary we can employ well-known tools in order to gain further intuition regarding orbital phenomena and chaotic imprints. A typical example is the Poincar\'e surface of section which is constructed by successive intersections of geodesics, with varying initial conditions, on a surface of section (here, we choose the equatorial plane) with strictly positive (or strictly negative) direction of intersection. The structure of the Poincar\'e map can instantly reveal if pure chaos is present, through disorganized intersections, or indirect imprints of non-integrability with the appearance of resonant islands, that encapsulate stable periodic points \cite{Contopoulos_book}, and exist due to non-integrability, in accord to the Kolmogorov-Arnold-Moser (KAM) and Poincar\'e-Birkhoff theorems \cite{Arnold_1963,Moser:430015,Birkhoff:1913}. Since the ZV metric does not have a Carter-like constant these features are present in its orbital phase space \cite{Lukes-Gerakopoulos:2012qpc}.

\begin{figure}[t]
	\includegraphics[scale=0.31]{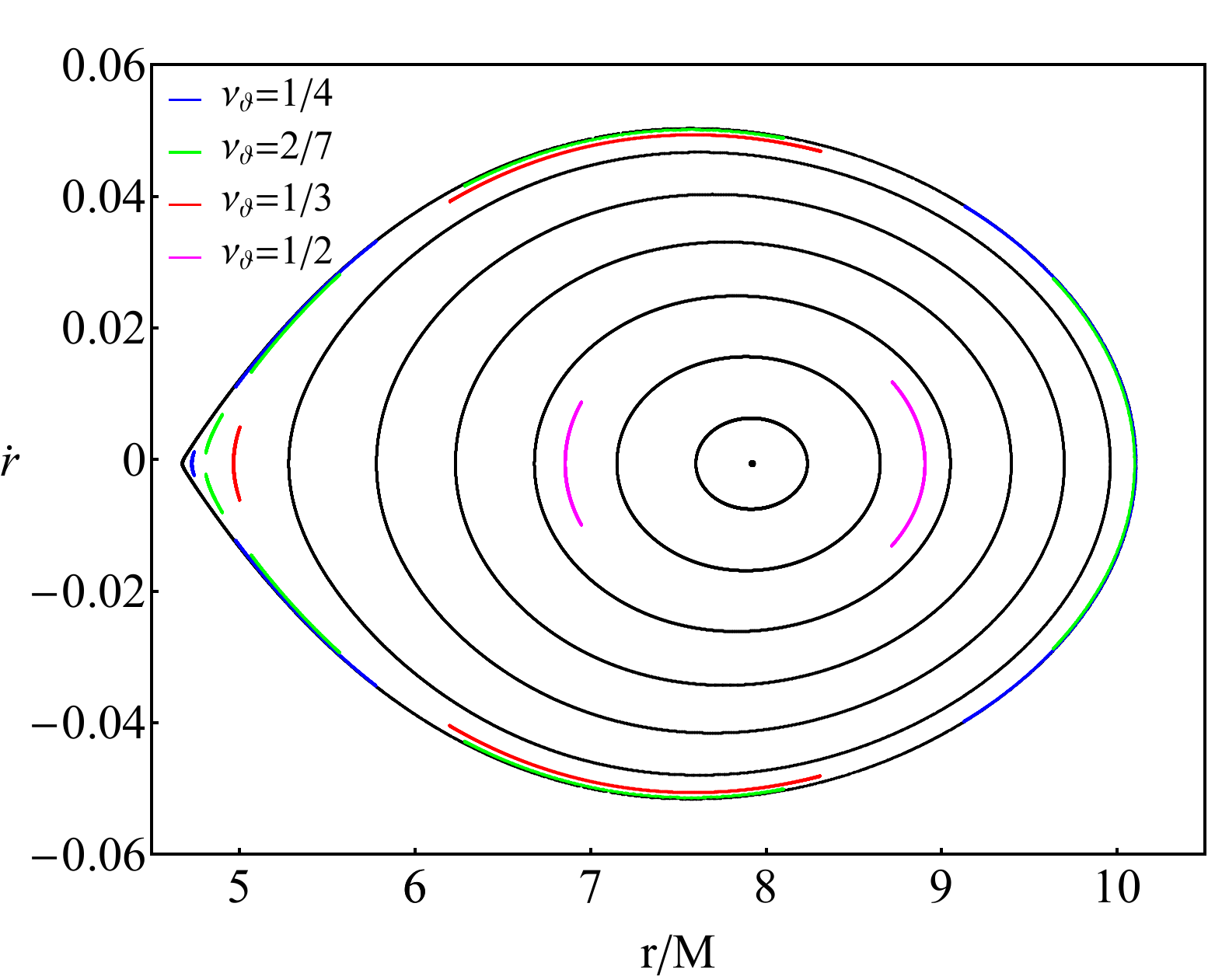}
	\caption{Poincar\'e map of a test-particle secondary with $\mu=1M_\odot$ orbiting around a ZV supermassive object with $M=10^6M_\odot$ and $\delta=1.5$. The radius of the edge of the CZV at $\dot{r}=0$ is $r=4.666M$. The secondary's conserved energy and angular momentum as $E/\mu=0.95$, $L_z/\mu=3M$, respectively. The initial condition for $r/M$ is varied smoothly while $\dot{r}=0$, $\theta=\pi/2$ and $\dot{\theta}$ is defined by the constraint equation \eqref{constraint}. Black curves that surround the central point of the map designate intersections through the equatorial plane of generic orbits, while colored curves designate intersections that belong to different resonant islands of stability.}\label{fig1}
\end{figure}

More specifically, indirect chaos, which we deal with in this work, is connected to the appearance of islands of stability that surround resonances in Poincar\'e maps. This is due to the fact that KAM curves at resonances disintegrate onto a set of stable and another set of unstable periodic points, instead of forming typical KAM curves. Stable periodic points are encapsulated by islands of stability (resonant KAM curves), where the rotation number is shared through all geodesics residing in the island, while unstable periodic points are sources of chaotic orbits that shield the islands with extremely thin chaotic layers of disordered intersections that are practically indistinguishable in most nonintegrable cases of non-Kerr BHs. Direct or pure chaotic orbits, on the other hand, occupy a much larger and clearly distinguishable volume of phase space in Poincar\'e maps and correspond again to disordered intersections on a Poincar\'e map (see e.g. Fig. 7 in Ref. \cite{Lukes-Gerakopoulos:2010ipp} or Figs. 9-12 in Ref. \cite{Lukes-Gerakopoulos:2012qpc}). In any case, pure chaos, or even thin chaotic layers around resonant islands, are not discernible in the Poincare map of Fig. \ref{fig1} due to the small deformation parameter we utilized.

Another tool to detect chaos is the rotation number. We calculate it by tracking the angle $\vartheta$ between two successive intersections on the Poincar\'e map relative to the fixed central point of the map which corresponds to a circular orbit that intersects the surface of section exactly at the same point. The rotation number is defined as the accumulation of many angles measured between consecutive intersections as \cite{Contopoulos_book}
\begin{equation}\label{rotation}
	\nu_\vartheta=\frac{1}{2 \pi N}\sum_{i=1}^{N}\vartheta_i,
\end{equation}
for which when $N\rightarrow\infty$ (with $N$ the number of angles measured), Eq. \eqref{rotation} converges to the radial and polar orbital frequency ratio $\nu_\vartheta=\omega_r/\omega_\theta$. Calculating consecutive rotation numbers for different geodesics, by smoothly varying one of the parameters or initial conditions of the system while keeping the rest fixed, leads to a rotation curve. Integrable systems demonstrate monotonous rotation curves. On the other hand, non-integrable systems possess transient plateaus with a non-zero width when geodesics occupy resonant islands. This designates a crucial aspect of resonant islands, that is when a geodesic is inside the island it shares the same rational ratio $\omega_r/\omega_\theta$ with the stable periodic point which leads to the plateau formation; a phenomenon that does not appear in integrable systems whatsoever, even though resonances still exist but occupy only a single point in phase space. Inflection points also appear when trajectories pass through the intersection of resonant islands where unstable periodic points reside and chaotic layers, that surround resonant islands, emanate. Nevertheless, by changing the initial conditions the orbit can be driven through the island and give rise to a typical plateau. 

Rotation curves are not only a tool that is used for geodesics but can also be employed in dissipative scenarios, where the rotation number evolves with respect to time. In the case of an EMRI, one can use selected timesteps of the inspiral as initial conditions for a geodesic evolution. Through the non-dissipative trajectory, the Poincar\'e map and eventually the rotation number of each timestep can be calculated in order to plot a series of rotation numbers as the EMRI evolves with time. The same attributes hold here as well, namely monotonous dissipative rotation curves for integrable systems and appearance of plateaus for non-integrable EMRIs. For more information regarding integrability and chaos in EMRIs see the following series of works and references therein \cite{Lukes-Gerakopoulos:2010ipp,Destounis:2020kss,Destounis:2021mqv,Destounis:2021rko}.

\begin{figure*}[t]
	\includegraphics[scale=0.345]{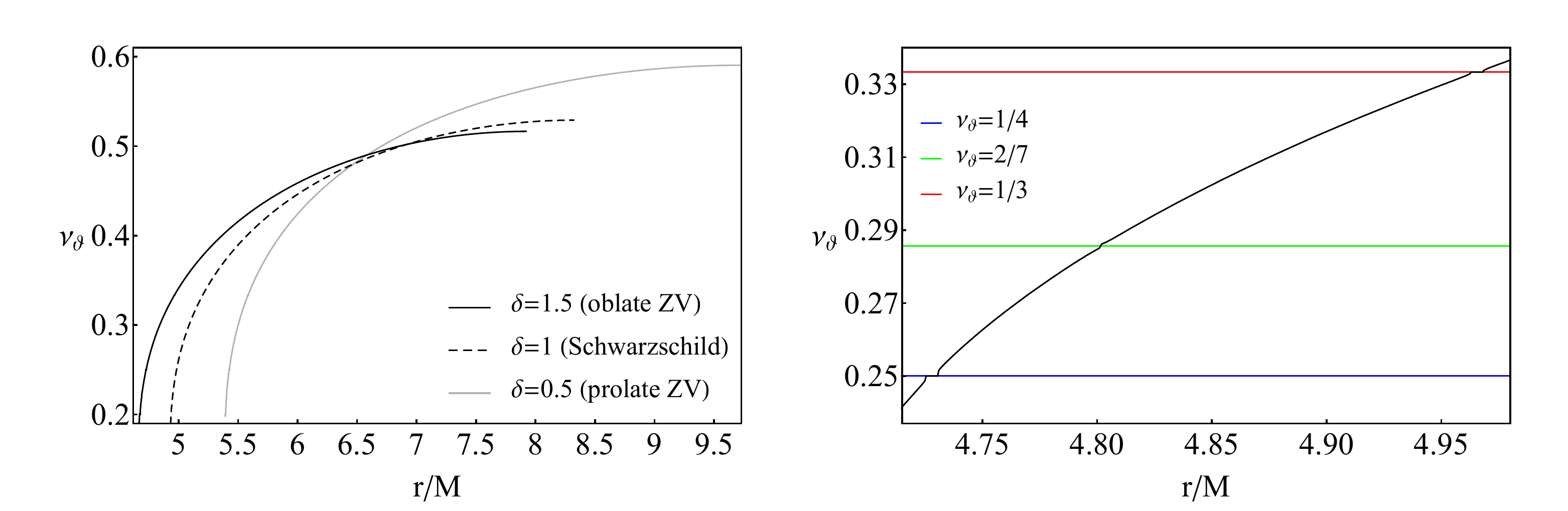}
	\caption{\emph{Left:} Rotation curves for geodesics of a test particle with $\mu=1M_\odot$, $E/\mu=0.95$ and $L_z/\mu=3M$ around a ZV object with $M=10^6M_\odot$ with varying deformation $\delta$. The radius where the CZV ends and orbits are plunging at $\dot{r}=0$ is $r=4.666M$ for $\delta=1.5$, $r=4.93M$ for $\delta=1$ and $r=5.396M$ for $\delta=0.5$. For completeness, we present the rotation curve for a Schwarzschild BH with the same parameters as above and $\delta=1$. The initial condition for $r/M$ is varied smoothly while $\dot{r}=0$, $\theta=\pi/2$ and $\dot{\theta}$ is defined by the constraint equation \eqref{constraint}. \emph{Right:} Zoom into a region of interest for the $\delta=1.5$ case. The horizontal colored lines designate where resonances appear.}\label{fig2}
\end{figure*}

\begin{figure*}[t]
	\includegraphics[scale=0.345]{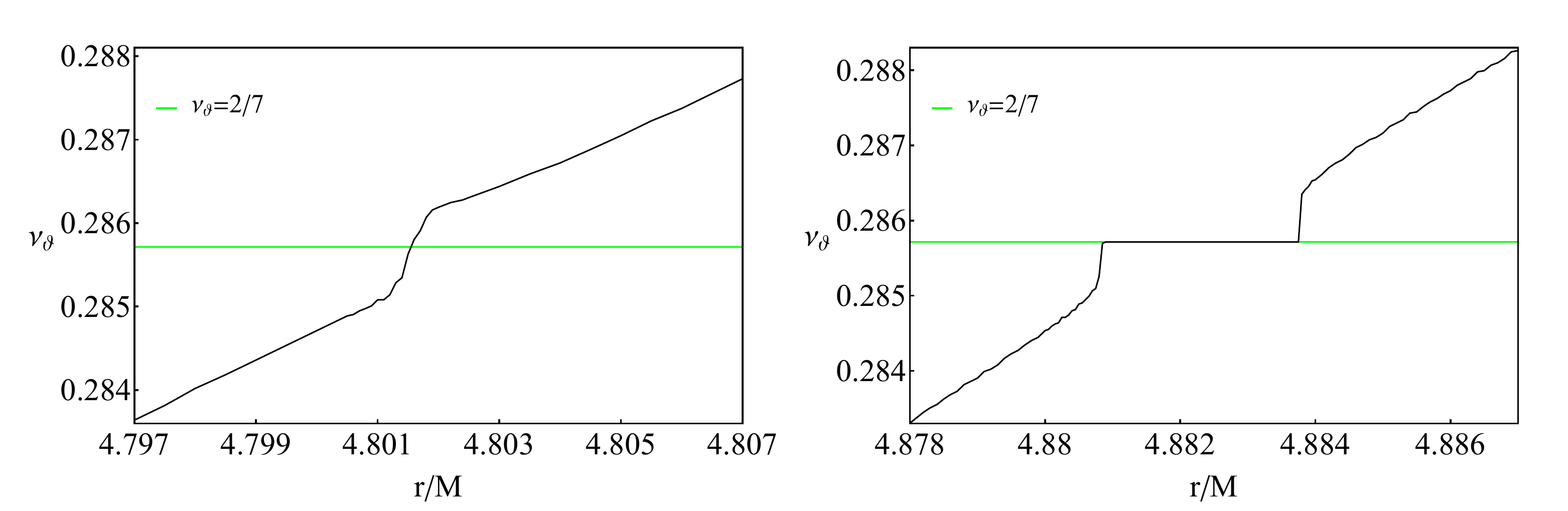}
	\caption{\emph{Left:} Zoom into the inflection point for the $2/7$-resonant island shown on the right subfigure of Fig. \ref{fig2}. initial conditions and parameter as the same as Fig. \ref{fig2}. \emph{Right:} Rotation curve for geodesics of a test particle with $\mu=1M_\odot$, $E/\mu=0.95$, $L_z/\mu=3M$ and $\dot{r}(0)=0.01$ around a ZV object with $M=10^6M_\odot$ with $\delta=1.5$.}\label{fig3}
\end{figure*}

\begin{figure*}[t]
	\includegraphics[scale=0.345]{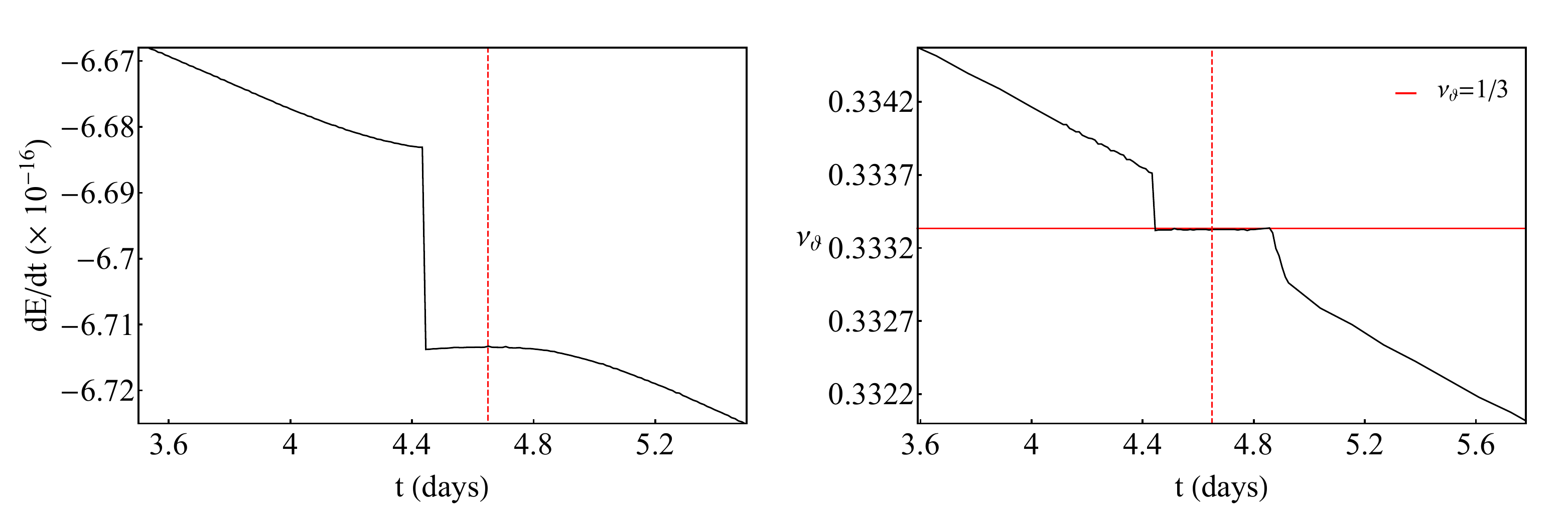}
	\caption{\emph{Left:} Evolution of energy flux $dE/dt$ for an inspiraling secondary with $\mu=1M_\odot$ and initial parameters $E_0/\mu=0.95$, $L_{z,0}/\mu=3M$, $r(0)=4.985M$, $\theta(0)=\pi/2$, $\dot{r}(0)=0$, where $\dot{\theta}(0)$ is defined by the constraint equation \eqref{constraint}, around a ZV primary with $M=10^6M_\odot$ and $\delta=1.5$. \emph{Right:} Evolution of the rotation number of the same inspiral as in the left panel. The horizontal solid red line designates the $1/3$-resonant island. For both plots, the vertical dashed red line corresponds the time when the secondary passes through the center of the island.}\label{fig4}
\end{figure*}

\begin{figure*}[t]
	\includegraphics[scale=0.345]{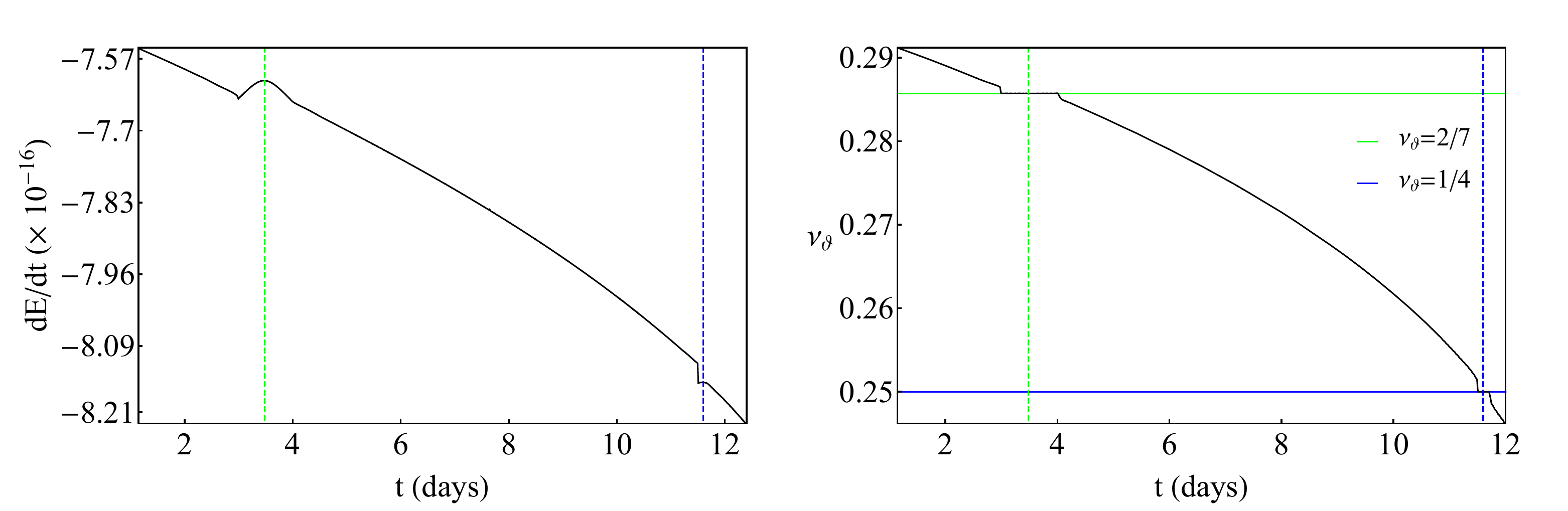}
	\caption{\emph{Left:} Evolution of energy flux $dE/dt$ for an inspiraling secondary with $\mu=1M_\odot$ and initial parameters $E_0/\mu=0.95$, $L_{z,0}/\mu=3M$, $r(0)=4.824M$, $\theta(0)=\pi/2$, $\dot{r}(0)=0$, where $\dot{\theta}(0)$ is defined by the constraint equation \eqref{constraint}, around a ZV primary with $M=10^6M_\odot$ and $\delta=1.5$. \emph{Right:} Evolution of the rotation number of the same inspiral as in the left panel. The horizontal solid colored lines designates the $2/7$ (green) and $1/4$-resonant islands (blue). For both plots, the vertical dashed colored lines corresponds the time when the secondary passes through the center of the corresponding islands.}\label{fig5}
\end{figure*}

\section{Geodesic and inspiral evolution}\label{geo_insp}

By solving the coupled $r,\,\theta$ second-order ordinary differential equations, together with the first-order decoupled equations for $t$ and $\phi$ from Eqs. \eqref{ELz}, we obtain bound orbits that reside inside CZVs and never plunge nor escape to infinity. To check the precision of our geodesics we evolve the constraint equation \eqref{constraint} for $\sim10^4$ revolutions and find that it is satisfied within one part in $10^{8}-10^{10}$ depending on the initial conditions and deformation of spacetime.

To guarantee numerical accuracy for inspirals, we calculate the $4$-velocity in each update of the fluxes and check its conservation along a geodesic evolution with initial conditions the energy, $z$-component of angular momentum, position and velocity at every update timestep. For all simulations presented hereafter, the constraint is satisfied to within a part in $\sim 10^8$ for the first $10^4$ crossings through the equatorial plane.

As a first step, we reproduce some qualitative features of the ZV metric, namely its non-integrability \cite{Lukes-Gerakopoulos:2012qpc}, which will be later used to choose  initial conditions in order to evolve EMRIs. In Fig. \ref{fig1} we show the Poincar\'e map for a ZV central object with $\delta=1.5$, meaning that the we utilized an oblate deformation with respect to Schwarzschild. We first observe a central point on the map. Since the map captures intersections of geodesics, the central point designates an orbit that is circular and always cuts the surface of section at the same $(r,\dot{r})$ position. Around the center, various black curves appear (known as KAM curves) that are formed through successive intersections of generic orbits with varying initial position $r(0)/M$. A Schwarzschild BH (or in general any spacetime with integrable geodesics) would exhibit a Poincar\'e map with KAM curves that only surround the central point of the map. The fact that our object does not have a fourth constant of motion (non-integrable), leads to the formation of nested islands around stable resonant points (see \cite{Destounis:2020kss} for a zoom into the encapsulated structure of these islands). 

\begin{figure*}[t]
	\includegraphics[scale=0.345]{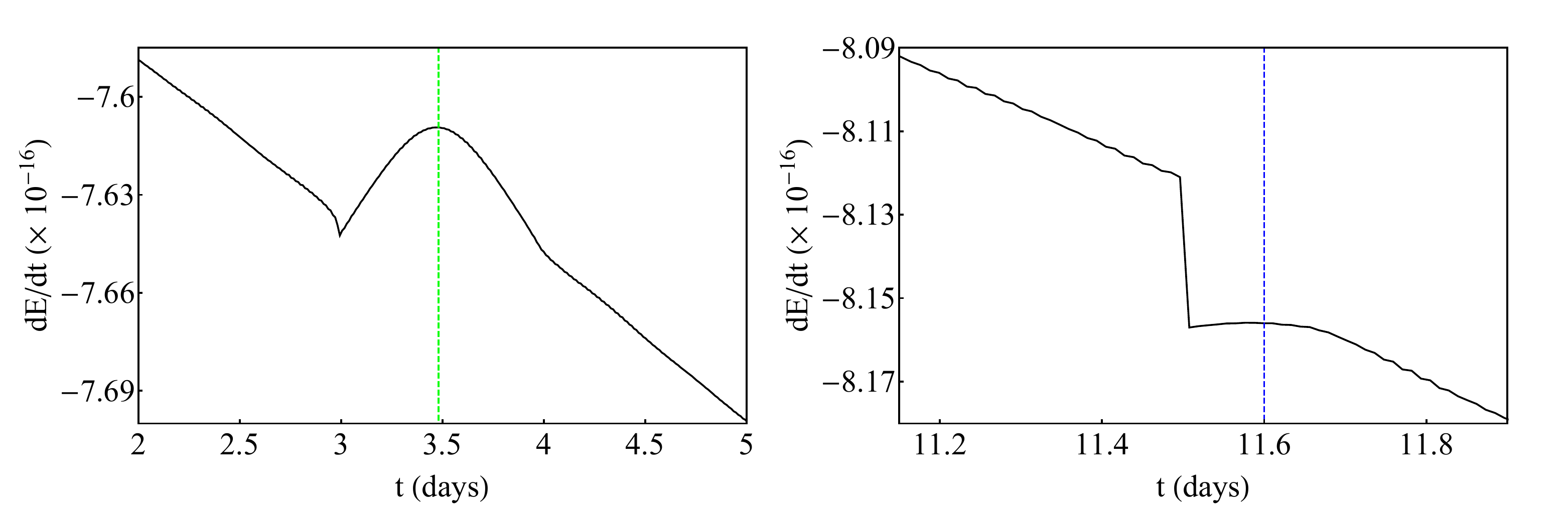}
	\caption{\emph{Left:} Zoom into the evolution of energy flux $dE/dt$ from Fig. \ref{fig5} through the $2/7$-resonant island. \emph{Right:} Same as the left panel for the $1/4$-resonant island crossing.}\label{fig6}
\end{figure*}

With a pedantic search on the available parameter space, we can easily spot four resonant islands with different multiplicity, which we designate in Fig. \ref{fig1} with colored curves. Notice how those do not surround the central point, but rather the stable points in phase space where the true resonant orbits emanate, and that the multiplicity defined by the denominator of the resonance (the longitudinal oscillations through the equatorial plane) corresponds to the number of islands. It is noteworthy that the resonances appearing here, besides the $1/2$-resonant island, are not those that strongly affect an inspiral, especially when assessing their impact in EMRI parameter estimation \cite{Flanagan:2010cd,Speri:2021psr,Gupta:2022fbe}, yet we will later see that they can also significantly contribute to it if the spacetime is non-integrable, mainly because they accumulate before plunge.

The rotation curves presented in Fig. \ref{fig2} promote the previous discussion perfectly. Various plateaus and inflections appear right where we expect the rotation number to have a rational ratio. The plateaus clearly designate that any geodesic residing in a resonant island shares the same rational ratio of orbital frequencies with the center of the island, where stable periodic orbits emanate. Even when an inflection point shows up at the rotation curve, meaning that the geodesic lies between the tips of two resonant islands where unstable periodic orbits exist, with a certain change in the initial velocity $\dot{r}(0)$, we can access the island as shown in Fig. \ref{fig3}. For completeness, we have considered both oblate ($\delta=1.5$) and prolate ($\delta=0.5$) deformations\footnote{In Fig. \ref{fig2} we also present a typical rotation curve for Schwarzschild spacetime.} and encounter similar effects. 

For the rest of the discussion we will focus on oblate deformations for the following reasons: (i) oblate deformations are usually the ones enabling the strongest effects of non-integrability and chaos (see \cite{Gair:2007kr,Lukes-Gerakopoulos:2010ipp,Lukes-Gerakopoulos:2014dpa,Destounis:2020kss,Destounis:2021mqv,Destounis:2021rko,Lukes-Gerakopoulos2020}), (ii) as seen in Fig. \ref{fig2}, an oblate deformation drives the resonances closer to the central object thus we expect amplified chaotic effects and (iii) $\delta>1$ is an optimal choice to imitate BHs since for these cases the ZV object possesses a PS, an ISCO and produces similar shadows \cite{Abdikamalov:2019ztb}.

So far, the discussion involved zeroth-order approximations of EMRIs since there was no radiation loss. Turning on the fluxes (rates of change of orbital energy and angular momentum) \eqref{update_E} and \eqref{update_L} leads to a numerical integration that is more intricate but the results are very interesting since the secondary inspirals adiabatically towards the primary due to fluxes through GWs. Fig. \ref{fig4} presents a typical inspiral of a secondary that transverses the $1/3$-resonant island. The initial conditions are not fine tuned so what is presented here is a generic feature. For this inspiral, we have updated the fluxes $N_r=800$ times, every $T_r=200M$. Since the revolution period in this case is $\sim 90M$, the total evolution time of the EMRI is $t_\text{total}=1.6\times 10^5M\sim 1800$ revolutions (roughly $9$ days for the mass of the primary considered).

At a certain time, the fluxes of energy and angular momentum become more negative abruptly. This instant of time designates the entry of the secondary into the $1/3$ plateau of the island (see right panel in Fig. \ref{fig4}). Right after the entry, the rate of change of orbital energy reaches a local maximum, at a special instant of time which is shown with a vertical dashed line, beyond which it continues to decline due to the inspiral. This behavior is not at all peculiar, but rather has a true physical meaning. Regardless of the fact that the secondary transverses the island while sharing the same rotation number throughout it, the secondary only reaches the stable periodic point at the center of the island at a certain time designated with the vertical dashed line. This is when the orbit becomes exactly periodic. Periodic orbits are the closest to circular ones. Since circular orbits emit monochromatic radiation at a single frequency (twice the revolution frequency) their energy emission is minimized (thus the rates of change of energy and angular momentum are also minimized (in absolute value)). Since a resonance emits quasi-monochromatic radiation, at the time that the inspiral passes through the perfect resonance, the fluxes are also maximized. The exit from the island takes place when the flux value at the entry is met again and the flux curve changes gradient (most obviously shown later in Fig. \ref{fig6}). The secondary can spend $\sim 100$ cycles in perfect resonance which translates to roughly half a day or $\sim 5\%$ of the whole EMRI evolution. The picture is qualitatively the same for the flux of angular momentum.

Fig. \ref{fig5} presents an even more intriguing EMRI evolution around the ZV primary, that undergoes two consecutive island crossings during a single evolution, namely the $2/7$ and $1/4$ islands. Initial conditions are slightly (but not strongly) fine tuned so what is presented here is quasi-generic, in a sense that it can occur for a small but non-zero set of initial conditions. For this inspiral, we have updated the fluxes $N_r=1100$ times, every $T_r=200M$. Since the revolution period in this case is $\sim 80M$, the total evolution time of the EMRI is $t_\text{total}=2.2\times 10^5M\sim 2800$ revolutions (roughly $12$ days for the mass of the primary considered). The time that the secondary spends in the first island, in perfect resonance, is $\sim 200$ cycles which translates to roughly a day. After exiting the first island, the orbit enters shortly after a subsequent resonant island and occupies it for another $\sim 50$ cycles which translate to one fifth of a day. In total the inspirals experiences $250$ cycles of perfect resonance (without taking into account pre- and post-resonant effects) which correspond to $\sim 10\%$ of the whole evolution. Similar analyses for non-Kerr EMRIs with the same mass ratios used here ($\mu/M=10^{-6}$) experience $250$ cycles in the most prominent $2/3$-resonant island \cite{Destounis:2021mqv,Destounis:2021rko}, though due to the much slower inspiral (resonances appear further from the central object) this only corresponds to $\sim 3\%$ of the whole evolution spent in a single island. Practically, the fact that a multitude of islands gather close to the plunge gives us the ability to probe subdominant resonances consecutively, in a short period of time, and have an EMRI that experiences a significant fraction of its evolution in resonance. Zooming in on the left plot of Fig. \ref{fig5} (see Fig. \ref{fig6}), we observe a complete agreement with what is demonstrated in Fig. \ref{fig4}. Especially for the $2/7$-resonant island (left plot in Fig. \ref{fig6}), the initial condition chosen here allows for the secondary to remain in it for a substantial amount of time that leads to a significant maximization of the energy flux. This means that the orbit enters deep into the island and probes the stable periodic point for a significant amount of time till it exits. The initial condition may look special, since a typical crossing would not maximize fluxes considerably, nevertheless, plateaus on dissipative rotation curves are still generic for a wide range of initial conditions. In this particular subfigure, one can also locate not only the entry in the island (with the sudden drop of the energy flux discussed above) but also the exit where a change of the gradient occurs. The same practically happens in all fluxes when an EMRI transverses an island of stability but only very prominent and long plateaus can reveal evidence of not only the entry but also the exit from the resonance.

\section{Gravitational waves}

In this section we approximate the dominant GW emission of an inspiraling stellar-mass secondary around an oblate ZV supermassive compact object and search for imprints of non-integrability in the waveforms produced by such EMRIs when detected by a LISA-like interferometer. For the GW modeling, we use the quadrupole approximation described below.

\subsection{Quadrupole formula}

The radiative component of the metric perturbation introduced by the secondary at luminosity distance $d$ from the source $\boldmath{T}$ can be read at the transverse and traceless gauge as
\begin{equation}\label{metpert}
	h^\text{TT}_{ij}=\frac{2}{d}\frac{d^2 Q_{ij}}{dt^2},
\end{equation}
where $Q_{ij}$ is the symmetric and trace-free (STF) quadrupole tensor
\begin{equation}
	Q^{ij}=\left[\int x^i x^j T^{tt}(t,x^i) \,d^3 x\right]^\text{STF},
\end{equation}
with $t$ being the coordinate time measured at very large distances from the detector. The source term of the secondary (which is treated as a point particle through a delta function) is
\begin{equation}\label{Ttt}
	T^{tt}(t,x^i)=\mu \delta^{(3)}\left[x^i-Z^i(t)\right],
\end{equation}
where $\mathbf{Z}(t)=(x(t),y(t),z(t))$ the position vector in pseudo-Cartesian coordinates and 
\begin{align}
    x(t)&=r(t)\sin\theta(t)\cos\phi(t),\\
    y(t)&=r(t)\sin\theta(t)\sin\phi(t),\\
    z(t)&=r(t)\cos\theta(t),
\end{align} 
the trajectory components with respect to flat spherical coordinates, under the assumption that our space-borne detector is positioned at infinity. Even though we have identified the Schwarzschild-like coordinates $(r,\theta,\phi)$ of the secondary's trajectory with flat-space coordinates, known in the literature as the ``particle-on-a-string'' approximation, and we assume a finite luminosity distance $d$ from the source, such prescription is not strictly valid. Nevertheless, it has been found to work very well when generating EMRI waveforms in GR \cite{Babak:2006uv}.

GWs can be projected onto two polarizations, $+$ and $\times$, with the introduction of two unit vectors, $\boldsymbol{p}$ and $\boldsymbol{q}$, which are defined with respect to a third unit vector $\boldsymbol{n}$ that points from the source to the detector. The triplet of unit vectors $\left(\boldsymbol{p},\,\boldsymbol{q},\,\boldsymbol{n}\right)$ is chosen so that they form an orthonormal basis. The polarization tensor components read
\begin{equation}
	\epsilon_+^{ij}=p^i p^j-q^i q^j,\,\,\,\,\,\,\epsilon_\times^{ij}=p^i q^j+p^j q^i,
\end{equation}
and allow for the metric perturbation to be written as
\begin{equation}
	h^{ij}(t)=\epsilon_+^{ij}h_+(t)+\epsilon_\times^{ij}h_\times(t),
\end{equation}
with
\begin{equation}
	h_+(t)=\frac{1}{2}\epsilon_+^{ij}h_{ij}(t),\,\,\,\,\,\,\,h_\times(t)=\frac{1}{2}\epsilon_\times^{ij}h_{ij}(t).
\end{equation}
The GW polarization components can then be described in terms of the position, $Z^i(t)$, velocity, $v^i(t)=dZ^i/dt$, and acceleration $a^i(t)=d^2Z^i/dt^2$ vectors as \cite{Canizares:2012is}
\begin{equation}
	\label{GW_formula}
	h_{+,\times}(t)=\frac{2\mu}{d}\epsilon^{+,\times}_{ij}\left[a^i(t)Z^j(t)+v^i(t)v^j(t)\right].
\end{equation}
LISA's response to an incident GW is rather complicated and depends on the antennae response patterns (see \cite{Cutler:1997ta,Barack:2003fp} for the full equations). Here we assume a detector that lies at a luminosity distance $d$ with orientation $\boldsymbol{n}=(0,0,1)$ with respect to the source, and utilize
\begin{equation}\label{LISA_response}
	h_{\alpha}(t)=\frac{\sqrt{3}}{2}\left[F^{+}_{\alpha}(t)h_{+}(t)+F^{\times}_{\alpha}(t)h_{\times}(t)\right],
\end{equation}
where $\alpha=(I,II)$ is an index representing the different antenna pattern functions $F_\alpha^+$, $F_\alpha^\times$ which can be found in Refs. \cite{Apostolatos:1994,Cutler:1997ta,Barack:2003fp}. For phenomenological purposes we will use a single-channel approximation, that is set $\alpha=I$ in Eq. \eqref{LISA_response}, since it is enough to accommodate the fundamental parts of gravitational radiation emitted by the EMRI, i.e. its phasing.

\subsection{Gravitational-wave frequency evolution and cumulative glitches}

To comprehend how a ZV EMRI imprints non-integrable effects, such as plateaus, onto its emitted waveform, we calculate the GW in the Einstein-quadrupole approximation for the particular inspiral outlined in Sec. \ref{geo_insp} that crosses two consecutive islands, namely the $2/7$ and $1/4$-resonances. We focus solely on this example in order to demonstrate that even subdominant resonances affect EMRI evolution, when the spacetime presents chaotic features.

\begin{figure*}[t]
	\includegraphics[scale=0.385]{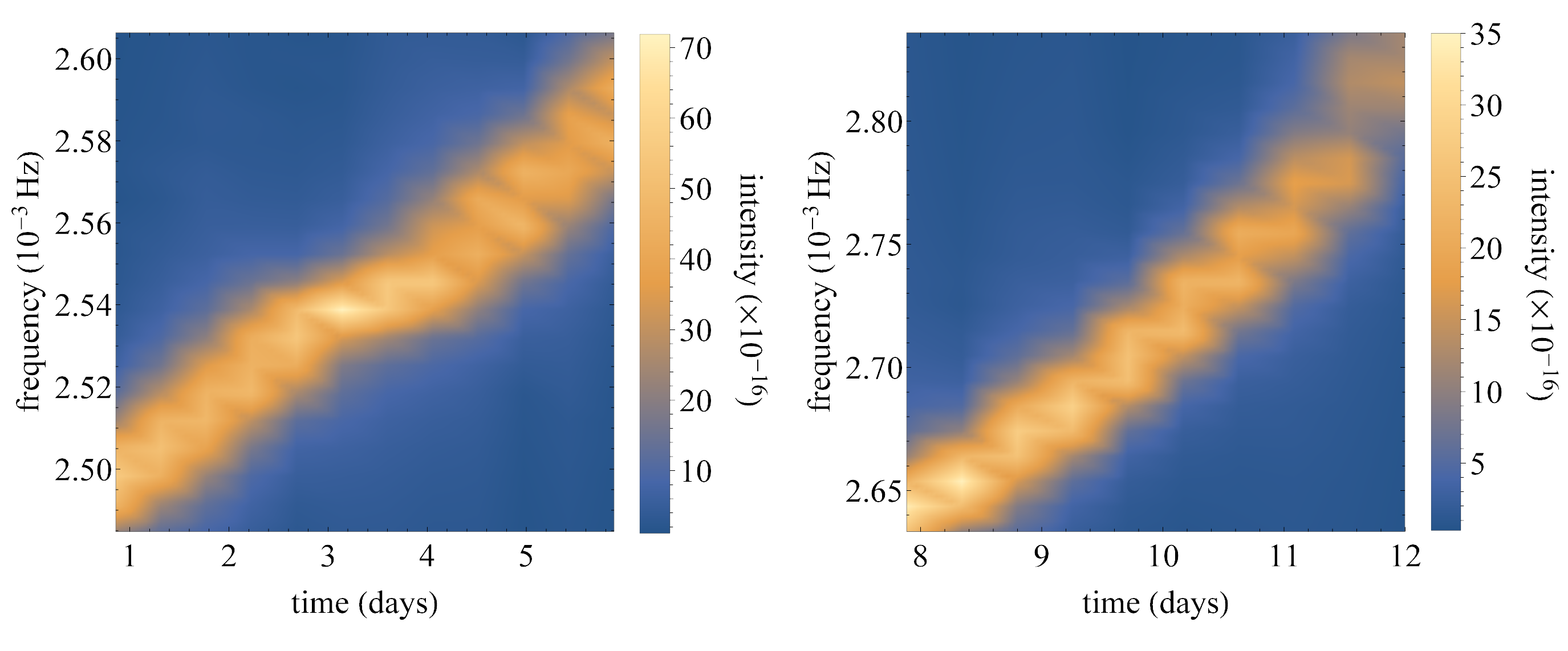}
	\caption{\emph{Left:} Frequency evolution of an EMRI through a consecutive crossing of resonant islands with parameters and initial conditions as in Figs. \ref{fig5} and \ref{fig6}. The approximate GWs are detected at luminosity distance $d=100\,\text{Mpc}$. The spectrogram presents the effect on a single harmonic of the waveform when it crosses the $2/7$-resonant island. \emph{Right:} Continuation of frequency evolution of the same EMRI with initial conditions as in the left panel. The spectrogram presents the effect on the waveform when it crosses the $1/4$-resonant island.}\label{fig7}
\end{figure*}

We perform a Fourier transform on the extracted waveform from the inspiral and plot a density spectrogram that displays the evolution of one of the harmonics with respect to time. We use the same methodology for the spectrogram as in \cite{Destounis:2021mqv,Destounis:2021rko}, i.e. we cut the waveform in time segments with a particular window size and offset and perform consecutive Fourier transforms, in order to overcome the uncertainty between frequency and time resolution. Here, since the inspiral is evolving fast and the island crossings have different timescales, we use distinct window size and offsets for the chunks of evolution around the two resonant islands. This is necessary for demonstration purposes since the first resonance is crossed much slower than the second one.

In Fig. \ref{fig7} we show the spectrogram of the GW extracted from the EMRI that successively transverses two resonant islands of stability. In both cases, when the island is met the waveform frequency evolution loses monotonicity. The resonant crossings manifest into the GW with either a plateau-like pattern (left plot in Fig. \ref{fig7}) or a rapid glitch\footnote{Unfortunately, the inspiral enters into a region of phase space right after the crossing of the 1/4-resonance where pure chaos is present and the Fourier peaks become continuous so we could not evolve the system for more time.} (right plot in Fig. \ref{fig7}). Both instances designate that the system is non-integrable, especially the encounter of the first island that shares a striking similarity with plateaus appearing in rotation curves. However, this is an outcome of short-term occupancy in the islands and altogether a smaller deformation of spacetime, with respect to those presented in non-Kerr EMRIs \cite{Destounis:2021mqv,Destounis:2021rko}. If the inspiral is given more time in the island, the manifestations should resemble more those of discontinuous GW glitches. Yet, the fact that resonant island crossings are imprinted in the GW of a ZV EMRI, in a short period of time, can serve as a `smoking gun' for a non-integrable BH mimicker in the center of our galaxy, since these phenomena do not appear in Schwarzschild EMRIs that have monotonous rotation curves and spectrograms.

\section{Concluding remarks}

Probing the spacetime around supermassive compact objects with EMRIs is one of the prime targets for space-borne detectors like LISA. Compact objects are either spinning and/or surrounded by astrophysical environments, especially those residing in active galactic nuclei. Therefore, spherical symmetry is rather fragile and is usually broken. When there are not enough spacetime symmetries to guarantee the integrability of geodesics around these objects, then LISA may be able to detect particular phenomenological imprints of such manifestations like GW glitches around expected transient orbital resonances \cite{Destounis:2021mqv,Destounis:2021rko} that differ significantly from instrumental glitches \cite{Edwards:2020tlp}.

We showed that deformations from spherical symmetry which keep the spacetime static, described by the ZV geometry, have the potential to exhibit GW glitches when involved as primary objects in EMRIs, due to the non-integrability of test-particle dynamics. These phenomena not only appear for single resonant island crossings, such as those occurring in non-Kerr EMRIs, but also are present in a cumulative and short-timescale manner, where the secondary transverses two (and possibly more) subdominant resonant islands. Note though that this can also occur in non-Kerr EMRIs if long-lasting evolutions are to be performed, though the effect of subdominant resonances are usually suppressed \cite{Destounis:2021mqv,Destounis:2021rko}.

When supermassive compact objects, such as those residing in galactic centers, have multiple interpretations due to the plethora of objects that can cast similar shadows, GW astronomy with LISA can, in principle, distinguish between models that are integrable or not through the detection of glitches in gravitational waveforms. Synergies between shadow observations and space-borne detectors can, therefore, narrow down the parameter space of solutions describing supermassive objects in galactic centers, and in particular can emphatically decide if M87* and Sgr A* are described by Kerr and Schwarzschild geometries, respectively, unless the EMRI includes multiple \cite{Barausse:2006vt,Amaro-Seoane:2011rdr} or spinning \cite{Kiuchi:2004bv,Zelenka:2019nyp} secondaries.

Our analysis only deals with the phenomenological imprints of non-integrability. Nevertheless, if one wants to utilize the aforementioned phenomenology in practice, a consistent glitch modeling analysis for non-integrable EMRIs should be carried out, in a systematic way in order to understand, first, if these phenomena are clearly detectable with space interferometers, second, to which extent they affect parameter estimation, and third, to which degree these effects differ from standard transient resonances experienced by integrable EMRIs which are already sufficiently modeled with PN techniques \cite{Speri:2021psr,Gupta:2022fbe} and gravitational self-force \cite{Detweiler:2005kq,Barack:2009ux,Flanagan:2010cd,Flanagan:2012kg,Berry:2016bit}.

\section*{Declarations}
We the authors hereby declare that there are no competing interests of financial or personal nature. Ethical approval for this work is not applicable. All authors have contributed equally for the execution of calculations and the presentation of results, as well as the creation of this manuscript. The data that support the findings of this study are available from the corresponding author upon reasonable request. This work was supported by the DAAD program for the ``promotion of the exchange and scientific cooperation between Greece and Germany IKYDAAD 2022" (57628320). K.D. and K.D.K. are grateful for hospitality provided by the Section of Astrophysics, Astronomy, and Mechanics, Department of Physics of the  National and Kapodistrian University of Athens, Panepistimiopolis Zografos GR15783, Athens, Greece. K.D. acknowledges financial support provided under the European Union's H2020 ERC, Starting Grant agreement no.~DarkGRA--757480 and the MIUR PRIN and FARE programmes (GW-NEXT, CUP: B84I20000100001). 

\bibliography{ZV}

\end{document}